\documentclass[
    aps,
    pra,
    amsmath,amssymb,
    reprint,
    superscriptaddress
    ]{revtex4-1}
    \newcommand{\expect}[1]{
        \ensuremath{\langle{#1}\rangle}
    }

    \usepackage{graphicx}
    \usepackage{color}
    \usepackage{dcolumn}
    \usepackage{siunitx}
    \usepackage{bm}
    \usepackage{braket}
    \usepackage{here}
    \begin{document}
    
    
    \title[]{Generalization of the output of variational quantum eigensolver by parameter interpolation with low-depth ansatz}
    
    \author{Kosuke Mitarai}
    \email{mitarai@qc.ee.es.osaka-u.ac.jp}
    \affiliation{Graduate School of Engineering Science, Osaka University, 1-3 Machikaneyama, Toyonaka, Osaka 560-8531, Japan.}
    \affiliation{Qunasys Inc. High tech Hongo Building 1F, 5-25-18 Hongo, Bunkyo, Tokyo 113-0033, Japan}
    \author{Tennin Yan}
    \email{yan@qunasys.com}
    \affiliation{Qunasys Inc. High tech Hongo Building 1F, 5-25-18 Hongo, Bunkyo, Tokyo 113-0033, Japan}
    \author{Keisuke Fujii}
    \email{fujii.keisuke.2s@kyoto-u.ac.jp}
    \affiliation{Graduate School of Science, Kyoto University, Yoshida-Ushinomiya-cho, Sakyo-ku, Kyoto 606-8302, Japan.}
    \affiliation{JST, PRESTO, 4-1-8 Honcho, Kawaguchi, Saitama 332-0012, Japan}
    
    \date{\today}
    
    \begin{abstract}
    The variational quantum eigensolver (VQE) is an attracting possible application of near-term quantum computers.
    Originally, the aim of the VQE is to find a ground state for a given specific Hamiltonian.
    It is achieved by minimizing the expectation value of the Hamiltonian with respect to an ansatz state by tuning parameters \(\bm{\theta}\) on a quantum circuit which constructs the ansatz.
    Here we consider an extended problem of the VQE, namely, our objective in this work is to ``generalize'' the optimized output of the VQE just like machine learning.
    We aim to find ground states for a given set of Hamiltonians \(\{H(\bm{x})\}\), where \(\bm{x}\) is a parameter which specifies the quantum system under consideration, such as geometries of atoms of a molecule.
    Our approach is to train the circuit on the small number of \(\bm{x}\)'s.
    Specifically, we employ the interpolation of the optimal circuit parameter determined at different \(\bm{x}\)'s, assuming that the circuit parameter \(\bm{\theta}\) has simple dependency on a hidden parameter \(\bm{x}\) as \(\bm{\theta}(\bm{x})\).
    We show by numerical simulations that, using an ansatz which we call the Hamiltonian-alternating ansatz, the optimal circuit parameters can be interpolated to give near-optimal ground states in between the trained \(\bm{x}\)'s.
    The proposed method can greatly reduce, on a rough estimation by a few orders of magnitude, the time required to obtain ground states for different Hamiltonians by the VQE.
    Once generalized, the ansatz circuit can predict the ground state without optimizing the circuit parameter \(\bm{\theta}\) in a certain range of $\bm{x}$.
    \end{abstract}
    
    \pacs{Valid PACS appear here}
    \maketitle
    
    \section{Introduction}\label{sec:intro}
        Recent progress in the experimental realization of quantum computers is stimulating the research of their possible applications.
        Quantum computers which might realize in near-future are often called noisy intermediate-scale quantum (NISQ) devices \cite{Preskill2018}.
        Among possible applications of NISQ devices, variational quantum algorithms have recently attracted much attention \cite{Peruzzo2013, OMalley2016, Kandala2017, Otterbach2017, Havlicek2018}.
        In this variational approach, quantum computers have parameters which are optimized using classical algorithms with respect to a cost function, such as an expectation value of a Hamiltonian.
        To apply variational quantum algorithms to a problem, we divide the problem into two parts; one part which can be readily computed classically, and the other which is hard for classical computers but easy for quantum computers.
        
        The most popular two of variational algorithms are the variational quantum eigensolver \cite{Peruzzo2013,Bauer2016,Kandala2017} (VQE) for quantum chemistry and materials science, and quantum approximate optimization algorithm \cite{Farhi2014, Farhi2016, Otterbach2017} (QAOA) for combinatorial optimization problems.
        There are also many theoretical proposals which extend the hybrid framework to machine learning \cite{Mitarai2018, Huggins2018, Farhi2018, Schuld2018, Schuld2018a, Benedetti2018, Benedetti2018a, Fujii2016} along with the experimental demonstrations \cite{Havlicek2018, Negoro2018}.
        We refer to these approach as quantum circuit learning (QCL) following Ref. \cite{Mitarai2018}.
        In VQE or QAOA, a problem instance is encoded into an Hermitian operator \(H(\bm{x})\), where \(\bm{x}\) is an input parameter of the instance, such as the geometry of a molecule, external fields applied to a quantum system, or topology of a graph.
        Then we try to minimize an expectation value of \(H(\bm{x})\) with respect to an ansatz state \(\ket{\psi(\bm{\theta})}\), which is generated by a parameterized quantum circuit, by variating the parameter \(\bm{\theta}\).
        The optimized parameter \(\bm{\theta}^*\) gives a good approximate of the ground state of \(H(\bm{x})\).
        On the other hand, the QCL tries to minimize the cost function \(L\) which depends on multiple inputs \(\{\bm{x}_\alpha\}\) and multiple ansatz state \(\{\ket{\psi(\bm{x}_\alpha,\bm{\theta})}\}\).
        Specifically, in supervised learning, the cost function measures the difference between the teacher data \(\{y_j\}\) and the corresponding output with respect to the set of input \(\{\bm{x}_\alpha\}\).
        We can, therefore, construct a trained quantum circuit that gives outputs close to the teacher, by minimizing the cost function \cite{Mitarai2018}.
        The main difference between VQE/QAOA and QCL is that whether the output is ``generalized''.
        In the machine learning approach, the output is generalized in the sense that a trained circuit can predict the correct answers even when it is provided with a wide range of unknown inputs.
        In contrast, VQE and QAOA are specialized in solving one problem with a specific input.
        However, in many-body physics and chemistry, there is a high demand to know the changes of important properties of the system, such as the ground state energy or the order parameter, with respect to the system parameter \(\bm{x}\), such as the distance between the atoms or the magnitude of the applied magnetic field.
    
        In this work, we extend VQE to give us ``generalized'' outputs in the sense described above.
        Specifically, we aim to find a trained circuit which can output approximate ground states in a certain range of the input \(\bm{x}\).
        Classical machine learning approach using the Boltzmann machine has succeeded in learning the ground states of quantum many-body systems \cite{Carleo2017}, however, we expect that the quantum ansatz is more suited for the objective.
        We first propose the ansatz which, we claim from numerical simulation, are able to represent ground states with the small number of parameters in Sec. \ref{sec:theory}.
        The ansatz includes the adiabatic state preparation (ASP) in the limit of the infinite depth, which always outputs the ground state of Hamiltonian \(H(\bm{x})\) with the sufficiently large annealing time and it is irrespective of the parameter \(\bm{x}\). 
        However, the depth-limited ansatz does not allow us to achieve the robustness of the ASP with respect to the change in the parameter \(\bm{x}\), as shown in Appendix by numerical simulations.
        Instead, the interpolation between the optimal parameters \(\{\bm{\theta}^*(\bm{x}_\alpha)\}\) of the ansatz at different inputs \(\{\bm{x}_\alpha\}\) can be utilized to achieve the objective of generalization.
        Numerical simulations using Hamiltonians of hydrogen molecule, $\mathrm{H}_3$ molecule, and water molecule in Sec. \ref{sec:simulation} support this claim.
        Especially, for hydrogen and $\mathrm{H}_3$ molecule, we obtained the precision on the order of $10^{-5}$ Hartree in between the training points.
        
        The application of the proposed method is not limited to NISQ devices.
        In fault-tolerant quantum computation, which is a long-term goal for the experiments, the approximate ground state prepared by our method can then be used as an input to the phase estimation algorithm for a more accurate energy determination.
    
    \section{Methods}\label{sec:theory}
        \subsection{Variational quantum eigensolver}\label{sec:VQE_rev}
        Here we briefly review the algorithm called VQE \cite{Peruzzo2013}.
        The VQE aims to find a ground state of a Hamiltonian \(H(\bm{x})\) using the variational approach, where \(\bm{x}\) are parameters that specify a quantum system such as a molecule.
        For an electronic problem with two-body interactions, Hamiltonians have the following form,
        \begin{equation}\label{eq:fermionic_hamiltonian}
            H(\bm{x}) = \sum_{ij} h_{ij}(\bm{x})c_i^\dagger c_j + \sum_{ijkl} h_{ijkl}(\bm{x})c_i^\dagger c_j^\dagger c_k c_l,
        \end{equation}
        where \(c_i^\dagger\) and \(c_i\) are fermionic creation and annihilation operators, and \(h_{ij}(\bm{x})\) and \(h_{ijkl}(\bm{x})\) are coefficients.
        When applying VQE to molecular problems, we usually start with Hamiltonians that are described in Hartree-Fock (HF) basis \cite{McArdle2018a}.
        We then map the fermionic Hamiltonian Eq. (\ref{eq:fermionic_hamiltonian}) to a qubit Hamiltonian using a conversion such as Jordan-Wigner or Bravyi-Kitaev transformations \cite{Kassal2010, Seeley2012}.
        After the transformation, Hamiltonians acting on an \(n\)-qubit system have the following form,
        \begin{equation}\label{eq:qubit_hamiltonian}
            H(\bm{x}) = \sum_{P \in \mathcal{P}_{diag}} h_{P}^{(HF)}(\bm{x}) P + \sum_{Q \in \mathcal{P}_{nondiag}} h_{Q}^{(cor)}(\bm{x}) Q,
        \end{equation}
        where \(\mathcal{P}_{diag} \subset \{I,Z\}^{\otimes n}\) is a set of Pauli products which only contains Pauli \(Z\)'s and are diagonal in the computational basis, and \(\mathcal{P}_{nondiag} \subset \{I,X,Y,Z\}^{\otimes n}\setminus \mathcal{P}_{diag}\) is a set of Pauli products which are not diagonal in the computational basis.
        \(h_{P}^{(HF)}(\bm{x})\) and \(h_{Q}^{(cor)}(\bm{x})\) are real-valued coefficients.
        The ground state energy calculated by the HF approximation only depends on the first sum of Eq. (\ref{eq:qubit_hamiltonian}).
        The second sum of Eq. (\ref{eq:qubit_hamiltonian}) determines the correlation energy, which is discarded in the HF approximation.
        For later convenience, we define the HF Hamiltonian as \(H_{HF} = \sum_{P \in \mathcal{P}_{diag}} h_{P}^{(HF)}(\bm{x}) P\), and the correlation Hamiltonian as \(H_{cor} = \sum_{Q \in \mathcal{P}_{nondiag}} h_{Q}^{(cor)}(\bm{x}) Q\).
        
        To find a ground state, we construct a specific ansatz \(\ket{\psi(\bm{\theta})}\) that depends on the variational parameter \(\bm{\theta}\).
        It is usually created by applying a parameterized unitary gate \(U(\bm{\theta})\) to an initialized state \(\ket{0}^{\otimes n}\).
        There are several theoretical proposals addressing the form of \(U(\bm{\theta})\) \cite{Wecker2015, McClean2016, Dallaire-Demers2018}.
        
        \subsection{Variational ansatz}\label{sec:extVQE}
        We extend the VQE so that an optimized quantum circuit gives us a generalized output with respect to the parameter \(\bm{x}\), which specifies Hamiltonians \(H(\bm{x})\), that is, we aim to construct a quantum circuit that outputs ground states of \(\{H(\bm{x})\}\).
        When we want to find a ground state and its energy of each Hamiltonian \(H(\bm{x})\), the ansatz, in general, can be an arbitrary one that has enough power to represent ground states of each \(\{H(\bm{x})\}\) by optimizing the parameter \(\bm{\theta}\) of the ansatz at each \(\bm{x}\) independently.
        However, for the objective considered here, the ansatz must be constructed in a form that includes the parameter \(\bm{x}\) and can represent the ground states.
        
        Our idea is to use an ansatz which is inspired by the ASP \cite{Farhi2014}.
        The ASP finds a ground state of a Hamiltonian \(H(\bm{x})\) by implementing a time-dependent Hamiltonian,
        \begin{equation}\label{eq:fermionic_hamiltonian}
            H_{ann}(t) = A(t)H_0 + B(t)H(\bm{x}),
        \end{equation}
        where \(A(T) = B(0) = 0\), \(A(0) = B(T) = 1\) and \(H_0\) is a Hamiltonian of which ground state is trivial, for a duration \(T\).
        The output of the ASP is always the ground state of \(H(\bm{x})\) when \(T\) is sufficiently large.
        The ASP with the large \(T\) can be regarded as an ansatz circuit parameterized by \(H(\bm{x})\) itself that outputs the ground state of \(H(\bm{x})\) without any optimization, that is, it does not require any training.
        
\begin{figure}
            \includegraphics[width=\linewidth]{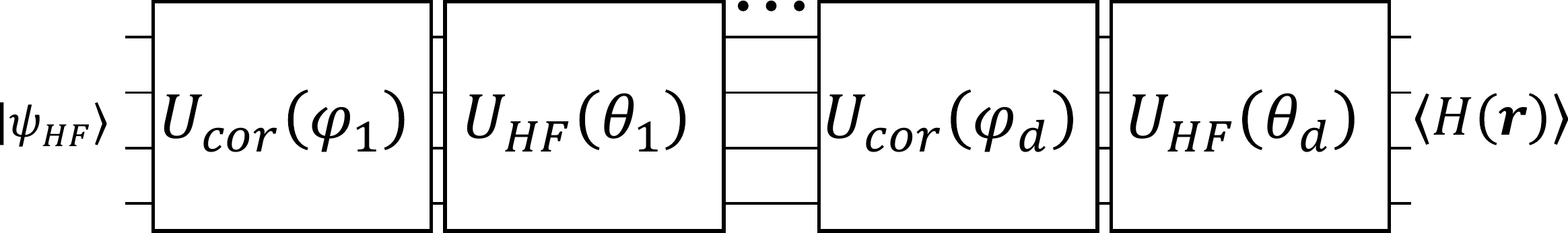}
            \caption{\label{fig:ansatz} Hamiltonian-alternating ansatz for generalized VQE.}
        \end{figure}

        The implementation of the ASP is unlikely on the NISQ devices, while it can be a nice starting point for the construction of the ansatz for the generalization considered here.
        We seek to resolve the objective considered here with what we call an Hamiltonian-alternating ansatz, 
        \begin{equation}\label{eq:ansatz}
            U(\bm{\theta}, \bm{\varphi}) = \prod_{k=1}^d U_{HF}(\theta_k)U_{cor}(\varphi_k).
        \end{equation}
        The quantum circuit corresponding to this is shown in Fig. \ref{fig:ansatz}.
        Hereafter we refer to the integer \(d\) as the depth of the ansatz circuit.
        \(U_{HF}(\theta_k)\) is a unitary operator generated by the HF Hamiltonian \(H_{HF}(\bm{x})\); \(U_{HF}(\theta_k) = e^{-i\theta_k H_{HF}(\bm{x})}\), and \(U_{cor}(\varphi_k)\) is a unitary operator generated by the non-diagonal part of the qubit Hamiltonian \(H_{cor}(\bm{x})\); \(U_{cor}(\varphi_k) = e^{-i\varphi_k H_{cor}(\bm{x})}\).
        We use the HF ground state \(\ket{\psi_{HF}}\) as the input to the circuit.
        This ansatz includes the ASP in the limit of \(d\to\infty\) and appropriate \(\bm{\theta}\) and \(\bm{\varphi}\) are chosen.
        The optimization of the circuit parameters \(\bm{\theta}\) and \(\bm{\varphi}\) at a certain \(d\) finds the shorter circuit than the ASP, although the optimized circuit with optimal \(\bm{\theta}\) and \(\bm{\varphi}\) can lose the robustness with respect to the change of \(\bm{x}\) of the ASP.
        The ansatzes having similar forms can be found in Ref. \cite{Farhi2014, Wecker2015}. Especially, in the first proposal of QAOA \cite{Farhi2014}, they pointed out the connection between the ansatz and quantum annealing.
    
        We need the Trotterization of \(U_{cor}(\varphi_k)\) for accurate execution of it.
        However, in this work, we simplify it to one step Trotter expansion considering the limited circuit depth of the NISQ devices;
        \begin{equation}\label{eq:non-trotterization}
            \tilde{U}_{cor}(\varphi_k) = \prod_{Q \in \mathcal{P}_{nondiag}} \exp (-i\varphi_k h_{Q}^{(cor)} Q),
        \end{equation}
        where the product is taken in an arbitrary order. 
        
        \subsection{Generalizing Variational Quantum Eigensolver}\label{sec:gen_VQE}
        To ``generalize'' the outputs of an optimized circuit, we first tried simultaneous minimization of \(H(\bm{x})\) at different \(\bm{x}_\alpha\)'s.
        The subscript \(\alpha\) denotes an index of points on which the quantum circuit is trained.
        In this approach, we set the cost function, which we aim to minimize by variating \(\bm{\theta}\) and \(\bm{\varphi}\), as
        \(\label{eq:cost}
            \sum_\alpha \bra{\psi(\bm{x}_\alpha, \bm{\theta}, \bm{\varphi})}H(\bm{x}_\alpha)\ket{\psi(\bm{x}_\alpha, \bm{\theta}, \bm{\varphi})}.
        \)
        Note that the ansatz state $\ket{\psi(\bm{x}_\alpha, \bm{\theta}, \bm{\varphi})}$ includes the parameter $\bm{x}$ since the ansatz is constructed from the Hamiltonian $H(\bm{x})$.
        Unfortunately, we found that this approach does not work well at small \(d\)'s, as shown in Appendix.
        With the low-depth ansatz, the optimization could not find the optimal \(\bm{\theta}\) and \(\bm{\varphi}\) that is robust to the change in the input parameter \(\bm{x}\).
        The reason for this may be that the compression of the ASP to the low-depth circuit depends on the parameter $\bm{x}$, and the depth of the circuit was traded-off with the robustness to the change in $\bm{x}$.
        The ASP is included in the limit of \(d\to\infty\) and in that case the ansatz must work, however, the depth of such a circuit is prohibitively large for the NISQ devices.
    
        To avoid the above issue and find a low-depth ansatz that returns a ground state irrespective to the change of the external parameter $\bm{x}$, we propose another approach to ``generalize'' with low-depth circuits.
        We seek a solution by assuming that the optimal \(\bm{\theta}\) and \(\bm{\varphi}\) themselves also depend on the input parameter \(\bm{x}\); \(\bm{\theta} \to \bm{\theta}(\bm{x})\) and \(\bm{\varphi} \to \bm{\varphi}(\bm{x})\), since we have found in the preliminary simulations that the optimal \(\bm{\theta}\) and \(\bm{\varphi}\) tended to have simple trends.
        To find these functions, we interpolate the optimal parameters determined at different \(\bm{x}\)'s.
        The procedure is as follows.
        First we optimize the parameter \(\bm{\theta}, \bm{\varphi}\) independently at different \(\bm{x}_\alpha\)'s.
        The resulting optimal parameters are denoted as \(\bm{\theta}^*(\bm{x}_\alpha)\) and \(\bm{\varphi}^*(\bm{x}_\alpha)\).
        Then we interpolate the points between \(\bm{\theta}^*(\bm{x}_\alpha)\) and \(\bm{\varphi}^*(\bm{x}_\alpha)\).
        The interpolation gives us the approximately optimal parameter function \(\bm{\theta}^*(\bm{x})\) and \(\bm{\varphi}^*(\bm{x})\) in between the training points \(\{\bm{x}_\alpha\}\).
        The ansatz in this approach can be written as:
        \begin{equation}\label{eq:ansatz}
            U(\bm{\theta}(\bm{x}), \bm{\varphi}(\bm{x})) = \prod_{k=1}^d U_{HF}(\theta_k(\bm{x}))\tilde{U}_{cor}(\varphi_k(\bm{x})).
        \end{equation}

        We note that the reduction in the number of parameters might be crucial for this approach.
    
        \section{Numerical simulation}\label{sec:simulation}
        For all simulations that we present in this section, the molecular Hamiltonian is calculated by OpenFermion \cite{McClean2017}, OpenFermion-Psi4 \cite{McClean2017}, and Psi4 \cite{Parrish2017}.
        We used the STO-3G minimal basis set and the Jordan-Wigner transformation for all calculations.
        Therefore the number of qubits in the simulated quantum circuit in this section were up to 14 for the water molecule in Sec. \ref{sec:water_sim}.

        At training points, the parameters were optimized using BFGS method \cite{Nocedal2006} provided in SciPy library.
        Quantum circuits were simulated with a variational quantum circuit simulator Qulacs \cite{Qulacs}.
        The gradients can be calculated using the method described in Ref. \cite{Mitarai2018,Guerreschi2017}.
        The starting points for the optimization were chosen randomly near the origin.
        More specifically, we sampled from the uniform distribution on \([0,10^{-2}]\) for initial values for each parameter \(\{\theta_k\}\), \(\{\varphi_k\}\).
        The optimization procedure is repeated 10 times starting from different initial parameters, and we picked the best one before the interpolation.
        After the optimal parameters are found, we used the quadratic interpolation.
        
        \subsection{Hydrogen molecule}\label{sec:H2_sim}
        First we consider the hydrogen molecule.
        In this simulation, the parameter \(\bm{x}\) of the Hamiltonian is the distance \(r\) between two hydrogen atoms.
        We search for the optimal parameter \(\bm{\theta}^*\), \(\bm{\phi}^*\) at \(\{r_\alpha\} = \) \(\{0.4, 0.6, 1.0, 1.4. 1.8, 2.2\}\) \(\mathrm{\AA}\).
        We found that the ansatz in Fig. \ref{fig:ansatz} with \(d=1\) can achieve the FCI energy almost to the machine precision, and therefore the all of the result in this section is for \(d=1\).
        
\begin{figure}
            \includegraphics[width=0.7\linewidth]{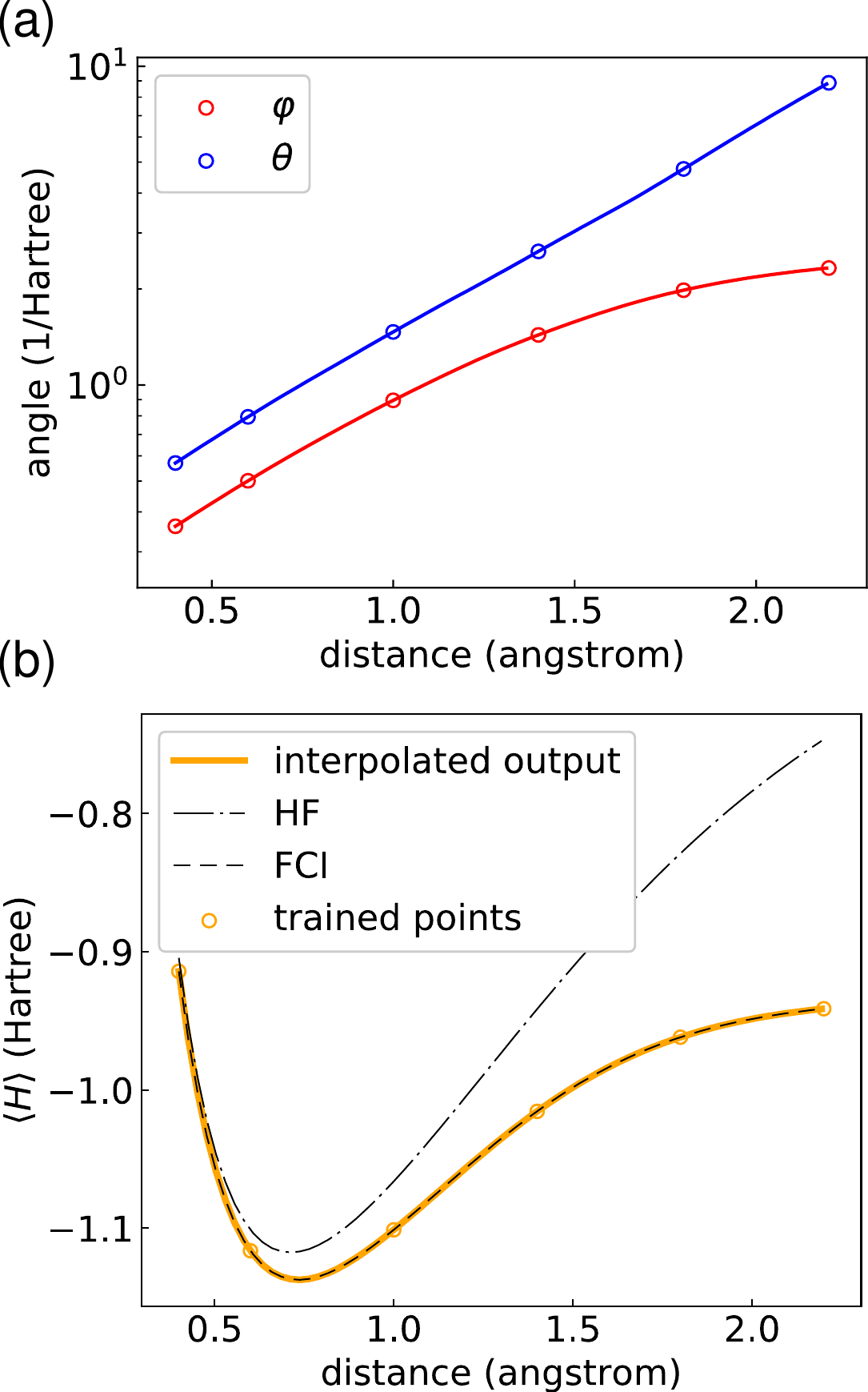}
            \caption{\label{fig:interpolation_H2} The result of numerical simulation for the hydrogen molecule with circuit depth \(d=1\). (a) Optimal parameters plotted against the interatomic distance \(r\). Circles: the best results among the optimizations started from 10 different initial values. Solid lines: drawn from quadratic interpolation between the markers. (b) Markers: optimized output at the training points. Solid line: output from the quantum circuit using the interpolated parameters. The HF and FCI energies are plotted for reference.}
        \end{figure}

        Fig. \ref{fig:interpolation_H2} (a) shows the dependency of optimal paramters.
        One can clearly see the trend in the optimal \(\theta\) and \(\varphi\) at the training points \(\{r_\alpha\}\) drawn by the markers.
        The solid lines in the figure are drawn from the interpolation.
        Using these interpoalted parameters, in Fig. \ref{fig:interpolation_H2} (b) we plot the output \(\expect{H(r, \bm{\theta}^*(r), \bm{\varphi}^*(r))}\) from the optimized quantum circuit.
        The output well matches the exact solution calculated by the FCI method.
        The error between the interpolated output and the FCI energy does not exceed \(6\times 10^{-5}\) Hartree.
        
        \subsection{Linearly aligned three hydrogen chain}\label{sec:linear_H3}
        Here we consider a somewhat artificial problem of finding ground state energies of the linearly aligned \(\text{H}_3\) molecule, to see if the proposed method works in bigger quantum systems.
        When we use the minimal basis set for this molecule, which is the case studied here, the system that we aim to solve can be regarded as a three-site Fermi-Hubbard model with periodic boundary conditions at half-filling.
        The coordinates of three hydrogen atoms are set to \(-r,~0,\) and \(r\), respectively, with \(r\) being the parameter \(\bm{x}\) which determines a Hamiltonian.
        Training points are set to \(\{r_\alpha\}=\) \(\{0.4, 0.6, 1.0, 1.4, 1.8, 2.2\}\) \(\mathrm{\AA}\).
    
\begin{figure}
            \includegraphics[width=0.7\linewidth]{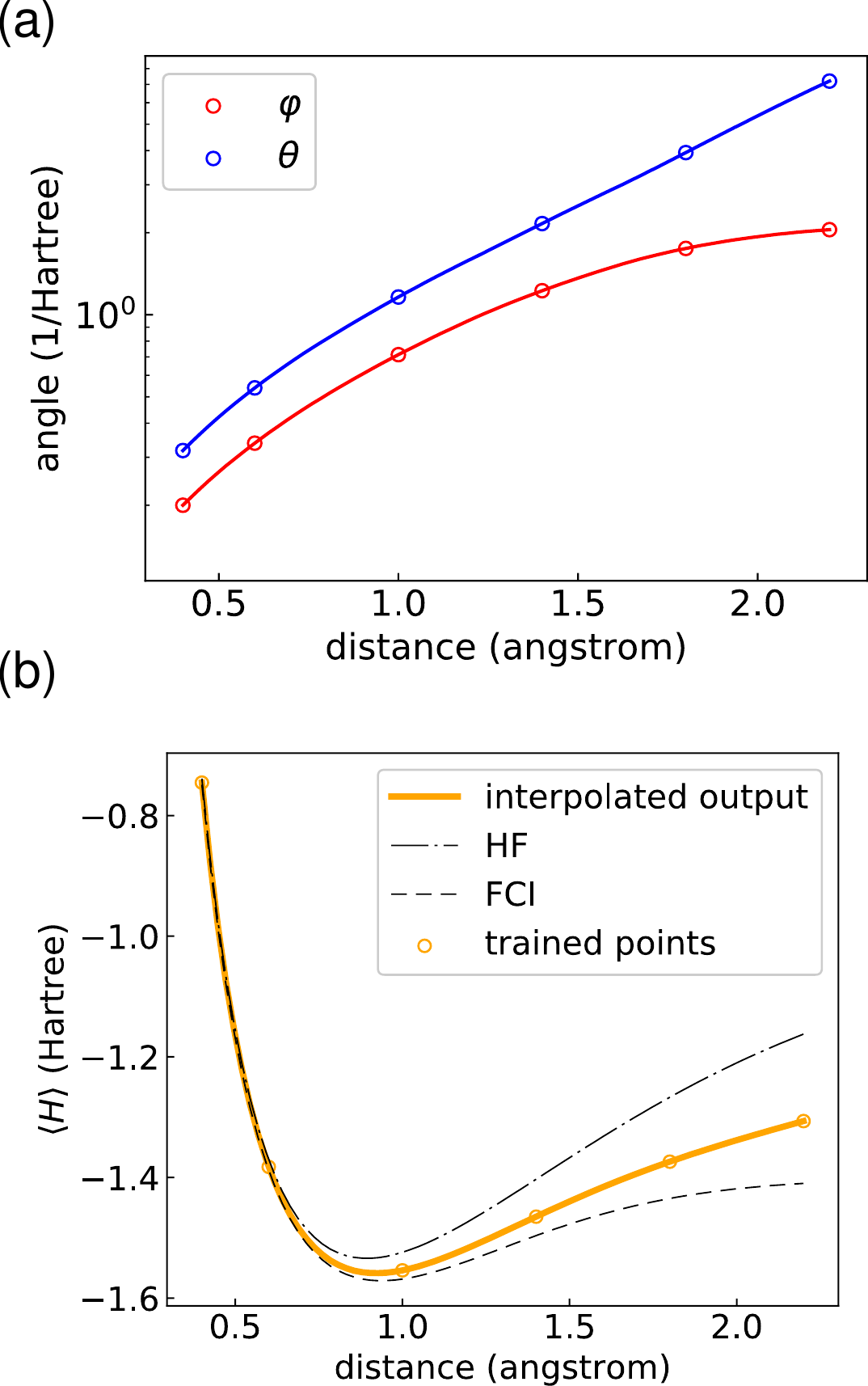}
            \caption{\label{fig:interpolation_linear_H3_d1} The result of numerical simulation for the linearly aligned \(H_3\) with circuit depth \(d=1\). (a) Optimal parameters plotted against the interatomic distance \(r\). Circles: the best results among the optimizations started from 10 different initial values. Solid lines: drawn from quadratic interpolation between the markers. (b) Markers: optimized output at the training points. Solid line: output from the quantum circuit using the interpolated parameters. The HF and FCI energies are plotted for reference.}
        \end{figure}

\begin{figure}
            \includegraphics[width=0.7\linewidth]{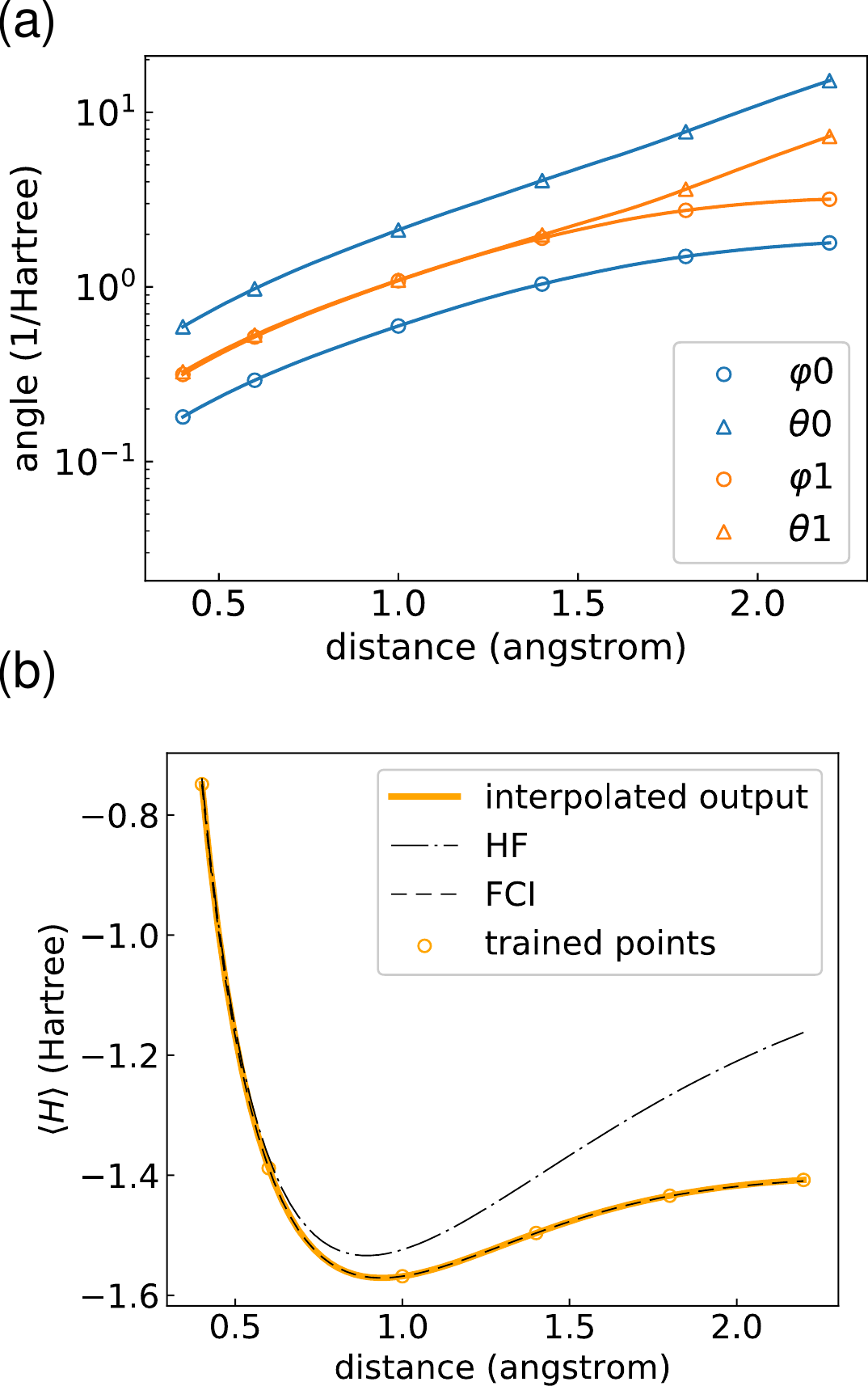}
            \caption{\label{fig:interpolation_linear_H3_d2} The result of numerical simulation for the linearly aligned \(\text{H}_3\) with circuit depth \(d=2\). (a) Optimal parameters plotted against the interatomic distance \(r\). Circles: the best results among the optimizations started from 10 different initial values. Solid lines: drawn from quadratic interpolation between the markers. (b) Markers: optimized output at the training points. Solid line: output from the quantum circuit using the interpolated parameters. The HF and FCI energies are plotted for reference.}
        \end{figure}

        The result for depth \(d=1\) and \(d=2\) are shown in Figs. \ref{fig:interpolation_linear_H3_d1} and \ref{fig:interpolation_linear_H3_d2}, respectively.
        From Fig. \ref{fig:interpolation_linear_H3_d1}, it is clear that the circuit with \(d=1\) is not capable of describing the ground state of the system since the output is not achieving the FCI energy.
        On the other hand, increasing the circuit depth to \(d=2\), the optimized output reaches the FCI energy as shown in Fig. \ref{fig:interpolation_linear_H3_d2}, indicating the circuit can generate the approximate ground state.
        It is notable that the number of parameter at \(d=2\) is only 4 while the dimension of Hilbert space under the search is \(\left(\begin{array}{c}6\\3\end{array}\right)=20\).
        The interpolation of the parameter also works nicely in this case, probably benefitting from the small number of parameters used in the ansatz.
        The maximum error between the FCI energy and the optimized output is \(2\times 10^{-3}\) Hartree at the training point \(r = 2.2 \mathrm{\AA}\)
    
        \subsection{Triangle \(\text{H}_3^+\) ion}\label{sec:triangle_H3_ion}
        We performed simulation on the triangle \(\text{H}_3^+\) ion.
        The hydrogen atoms are placed at \((0,0)\), \((r,0)\), and \((r/2,\sqrt{3}r/2)\), with \(r\) being the parameter which determines the Hamiltonian.
        Here we use \(\{r_\alpha\} = \{0.5, 1.0, 1.5, 2.0, 2.5\}\mathrm\AA\)
        We found, as same as in the case of the previous section, the output well converges to FCI energy at \(d=2\).
    
\begin{figure}
            \includegraphics[width=0.7\linewidth]{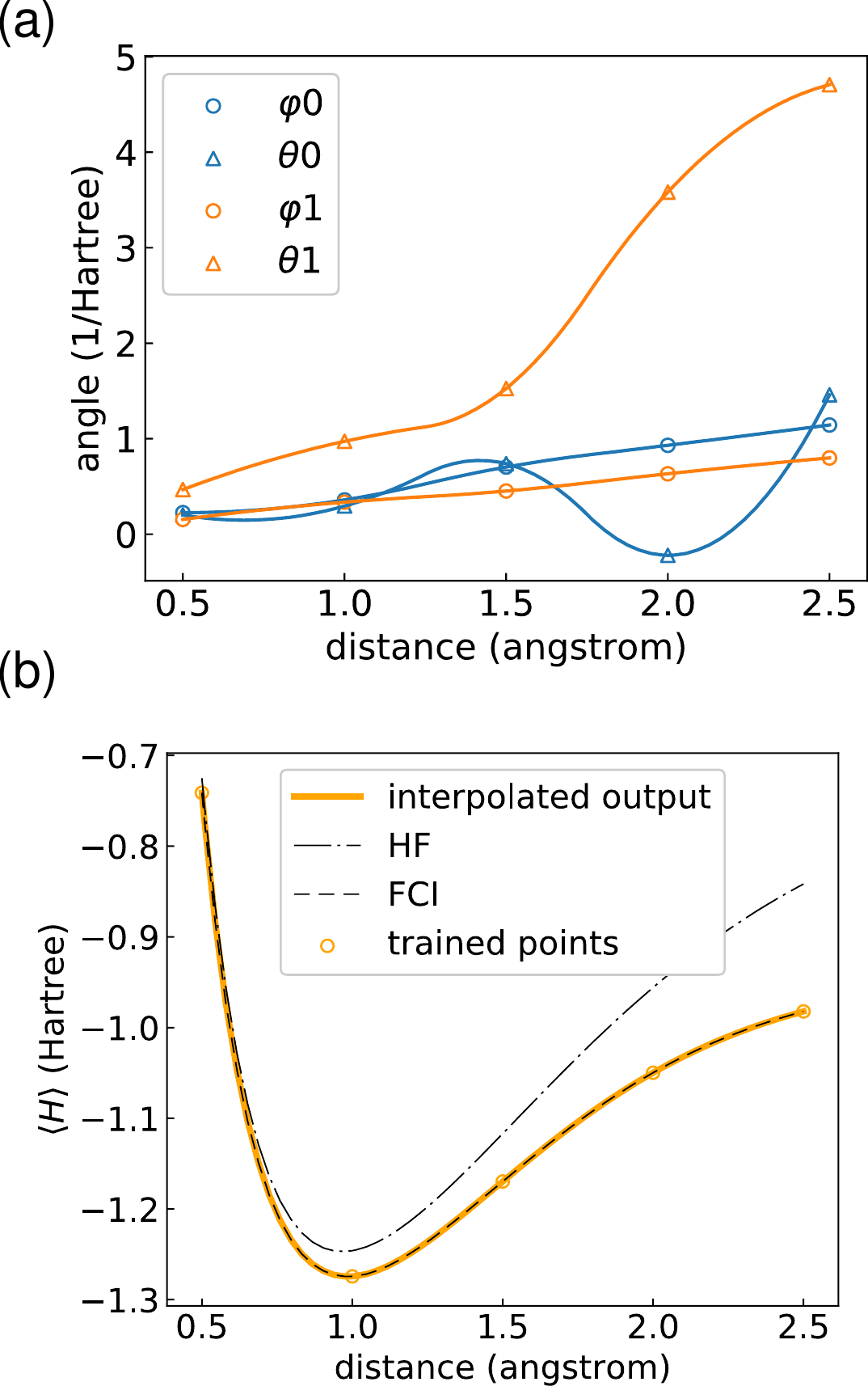}
            \caption{\label{fig:interpolation_triangle_H3_d2} The result of numerical simulation for triangle \(\text{H}_3^+\) with circuit depth \(d=2\). (a) Optimal parameters plotted against the interatomic distance \(r\). Circles: the best results among the optimizations started from 10 different initial values. Solid lines: drawn from quadratic interpolation between the markers. (b) Markers: optimized output at the training points. Solid line: output from the quantum circuit using the interpolated parameters. The HF and FCI energies are plotted for reference.}
        \end{figure}

        Fig. \ref{fig:interpolation_triangle_H3_d2} shows the simulated result of \(d=2\).
        The optimal parameters do not appear to have the simple trend as in the previous two examples, nevertheless, the interpolation approach works well in this case too.
        The error between the FCI energy and the output does not exceed \(2\times 10^{-5}\) Hartree.
    
        \subsection{Water molecule}\label{sec:water_sim}
        Here we simulate the water molecule \(\text{H}_2\text{O}\).
        In this simulation, we fix the bond length \(r\) between H and O to a constant (\(r=0.96\mathrm{\AA}\)) and set the angle \(\beta\) of H-O-H bonding as the parameter of the Hamiltonian;
        the oxygen atom was placed at \((0,0)\) and the two hydrogen atoms were placed at \((r,0)\) and \((r\cos\beta,r\sin\beta)\).
        The optimization was performed at \(\{\beta_\alpha\} = \{54, 72, 90, 108, 126, 154\}~\text{deg}\).
    
\begin{figure}
            \includegraphics[width=0.7\linewidth]{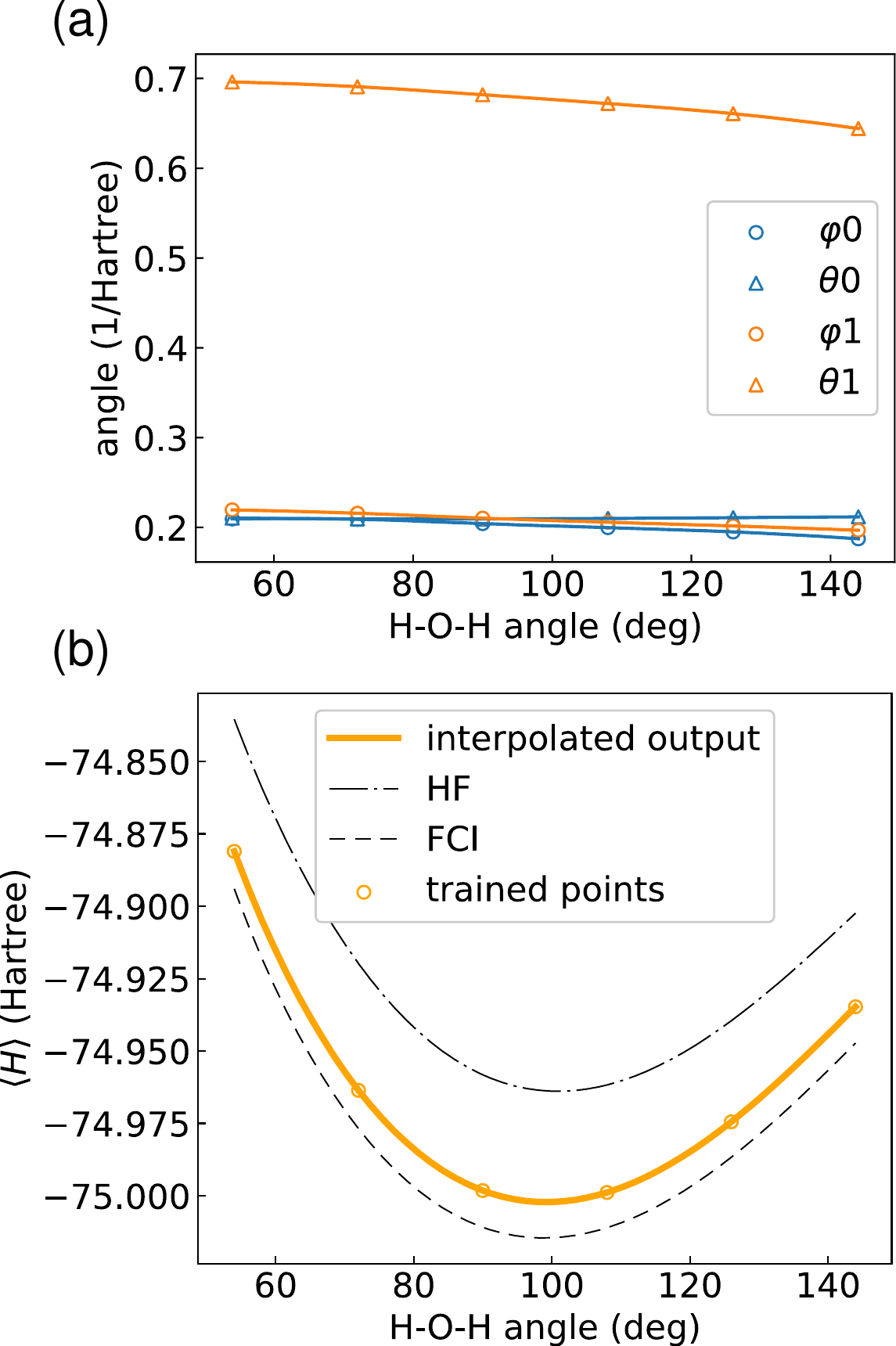}
            \caption{\label{fig:interpolation_H2O_d2} The result of numerical simulation for \(\text{H}_2\text{O}\) with circuit depth \(d=2\). (a) Optimal parameters plotted against the angle of H-O-H bonding \(\beta\). Circles: the best results among the optimizations started from 10 different initial values. Solid lines: drawn from quadratic interpolation between the markers. (b) Markers: optimized output at the training points. Solid line: output from the quantum circuit using the interpolated parameters. The HF and FCI energies are plotted for reference.}
        \end{figure}

        The result for \(d=2\) are shown in Fig. \ref{fig:interpolation_H2O_d2}.
        In this case, the optimal parameters give the better approximation than the HF approximation but do not reach the FCI energy.
        It is due to the limited representation power at \(d=2\).
        From the energy plot in Fig. \ref{fig:interpolation_H2O_d2} (b) we conclude that the interpolation approach can be utilized even at the level of 14 qubits.
        However, we found that, at \(d=3\) and \(d=4\), the parameter tends to be trapped at a local minima due to a complicated landscape of the energy expectation value.
        In those cases, the parameter found by the optimization process did not have the simple trends as in the figures, therefore the interpolation fails to predict the ground state energy in between the training points.
        More sophisticated optimization algorithms, such as the imaginary time evolution \cite{McArdle2018}, or the usage of the optimal parameter at one training point as the initial parameter for another training point, can be a solution to this problem by finding the global minimum of the energy.
    
        \section{Discussion}
        In Ref. \cite{Kandala2017}, the experimental VQE of \(\text{H}_2\), \(\text{LiH}_2\), \(\text{BeH}_2\) and the Heisenberg model took \(\approx 200\)-\(300\) iterations to converge.
        They employed the simultaneous perturbation stochastic approximation method for the optimization.
        They took \(10^3\) samples in the optimization procedure, resulting \(10^5\) to \(10^6\) samplings in total at one \(r\), the interatomic distance.
        Thus, for example, to plot an energy landscape using \(10^2\) points of \(r\), the number of samples can go up to \(10^8\).
        Since, from numerical simulations, our approach only needs around \(10\) optimizations at different \(r\)'s for the same task, the requirement can be dramatically lowered to the tenth of the original approach.
        Note that the direct interpolation between the calculated energies \(\{\expect{H(\bm{x}_\alpha)}\}\) might be able to do the same task, but our approach can prepare useful quantum outputs, that is, the approximate ground states in a certain range of \(\bm{x}\), not only the classical output like the energy.
        The approximate ground state can provide us other observables such as electron density.
        Also, we have observed that the direct interpolation approach tends to give a poorer approximation than the interpolation of the parameter utilized in this work. 
    
        To conclude, we proposed an idea to ``generalize'' the output of VQE.
        The simple interpolation of the circuit parameters, not of the calculated energy, enables us to resolve the objective.
        We gave the numerical evidence for the effectiveness of the approach.
        If one needs more accuracy, the interpolated values of parameters can be used as the starting point of the optimization for VQE.
        Our proposal can greatly reduce the time required for obtaining the ground states for a set of Hamiltonian \(\{H(\bm{x})\}\) characterized by some parameter \(\bm{x}\), and therefore advances the practicability of VQE.

        \begin{acknowledgements}
            KM would like to thank Yasunari Suzuki for fruitful discussions.
            KF is supported by KAKENHI No.
            16H02211, JST PRESTO JPMJPR1668, JST ERATO
            JPMJER1601, and JST CREST JPMJCR1673.
        \end{acknowledgements}
    
        \appendix*
        \section{Simultaneous optimization}
        In this Appendix we show the result with simultaneous optimization approach which is mentioned in Section \ref{sec:gen_VQE}.
        Unfortunately, we find that this approach does not work well at small \(d\)'s that we simulated.
        The compression of the ASP to the low-depth circuit depends on the parameter $\bm{x}$, and the depth of the circuit might be traded-off with the robustness to the change in $\bm{x}$.
        This has led us to explore the approach described in the main text.
        For numerical simulations, same procedures and packages as the main text is utilized for following numerical simulations.

        \subsection{\(\text{H}_2\)}
        In this simulation, the parameter \(\bm{x}\) of the Hamiltonian is the distance \(r\) between two hydrogen atoms.
        
\begin{figure}
          \includegraphics[width=\linewidth]{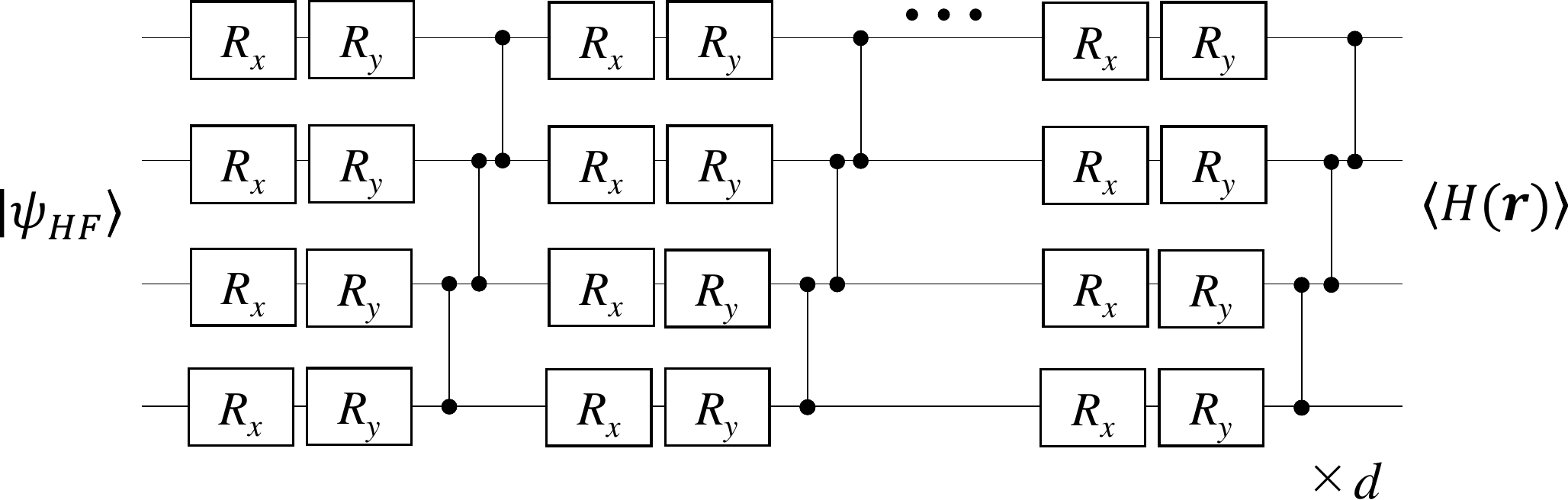}
          \caption{\label{fig:HE_ansatz} Hardware efficient circuit that is used for comparison. \(R_x\) and \(R_y\) are rotation gates around \(x\) and \(y\) axis respectively. Rotation angles of each rotation gate are independent parameters to be optimized.}
        \end{figure}

        We compare the result with our ansatz (Fig. \ref{fig:ansatz}) and that with so-called hardware efficient (HE) ansatz \cite{Kandala2017} (Fig. \ref{fig:HE_ansatz}).
        Parameters are optimized to minimize the cost Eq. (\ref{eq:cost}) with BFGS method, which is a gradient-based algorithm, using python library SciPy \cite{SciPy}.

        Fig. \ref{fig:H2_optimized} (a) and (b) show an example of the output from the optimized circuit with our ansatz and the hardware efficient ansatz with the same circuit depth \(d = 5\), respectively.
        We performed the optimization using molecular Hamiltonians at \(r = \) 0.4, 0.6, 1.0, and 1.4 \(\mathrm{\AA}\).
        In spite of the fact that the quantum circuit of the HE ansatz does not depend on Hamiltonians and always outputs the same quantum state, the energy obtained from the HE ansatz achieves quite close to FCI energy in a wide range.
        However, due to the lack of the information about the Hamiltonians to be minimized in the circuit, it performs poorly when compared with the Hamiltonian-alternating ansatz.
        This is apparent especially at the training points of \(r = \) 0.4 and 1.4 \(\mathrm{\AA}\).
        With the Hamiltonian-alternating ansatz, one can see that the trained points are actually optimized at the same time, and also that the optimized circuit gives us the energies that are close to the ones obtained from FCI calculations, even in the ranges between the trained points.
    
        Fig. \ref{fig:H2_optimized} (c) shows the depth dependency of our algorithm.
        We have conducted optimizations with the Hamiltonian-alternating ansatz with depths \(d =\) 1 to 9.
        Optimization was performed for 10 times for each depth starting from different initial parameters that are chosen randomly, and we picked an optimized circuit that gave us the lowest cost.
        One can see that with increasing depth of the ansatz, the trained points get close to FCI energies, as expected.
        On the other hand, increasing depth sometimes lowers the performance in between the training points.
    
        \subsection{Linearly aligned \(H_3\)}
        The coordinates of three hydrogen atoms are set to \(-r,~0,\) and \(r\), respectively, with \(r\) being the parameter \(\bm{x}\) which determines a Hamiltonian.
        We trained quantum circuits on the four values of the parameter \(r=\) 0.6, 1.0, 1.4, and 1.8 \(\mathrm{\AA}\), using the same method as the previous section.
    
\begin{figure}[H]
            \includegraphics[width=\linewidth]{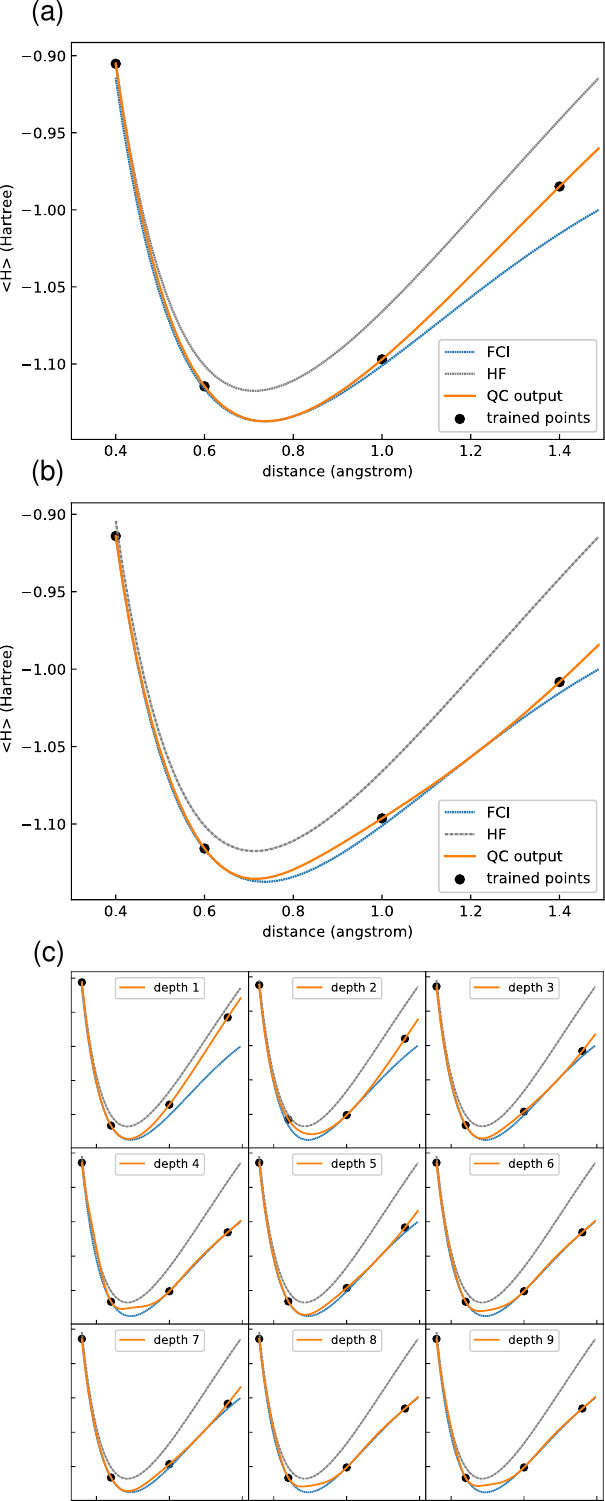}
            \caption{\label{fig:H2_optimized} (a), (b) Example of optimized outputs from the hardware efficient circuit (Fig. \ref{fig:HE_ansatz}) and the Hamiltonian-alternating circuit (Fig. \ref{fig:ansatz}), respectively. Depth \(d\) was set to 5, and the circuits are trained with the \(\text{H}_2\) molecular Hamiltonians of interatomic distances \(r=\) 0.4, 0.6, 1.0, and 1.4 \(\mathrm{\AA}\). The result from the Hartree-Fock (HF) calculations and the full configuration interaction (FCI), which is the exact ground state energy with the chosen basis set, are drawn for references. (c) Depth dependence of optimization result with the Hamiltonian-alternating ansatz.}
        \end{figure}

\begin{figure}[H]
            \includegraphics[width=\linewidth]{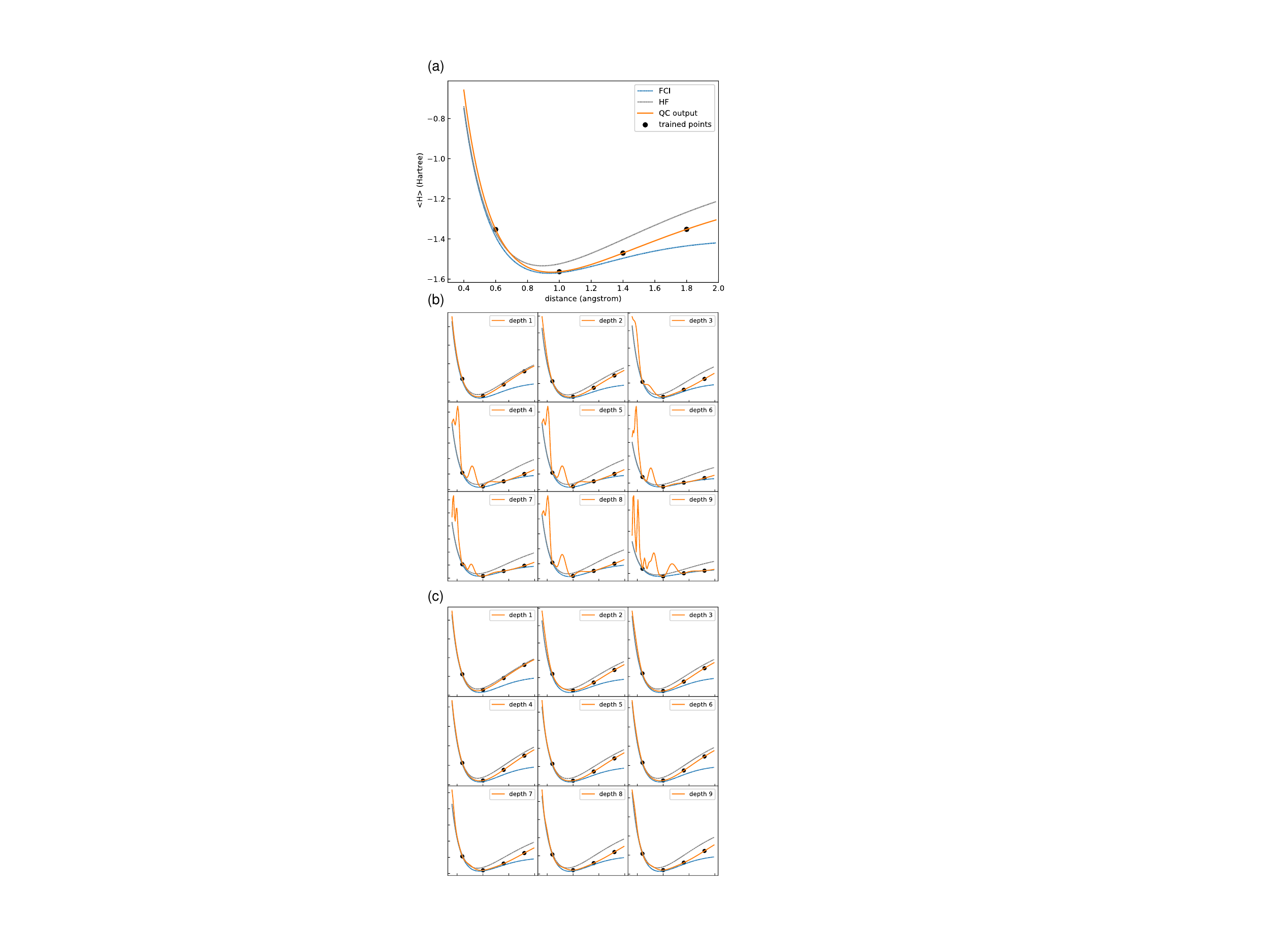}
            \caption{\label{fig:linear_H3_optimized} Simulation results for linearly aligned \(\text{H}_3\) molecular Hamiltonian. (a) Optimized output from the hardware efficient circuit for comparison. Depth \(d\) is set to 5. (b) Depth dependency of the optimization result with the Hamiltonian-alternating ansatz. (c) The optimization result with a threshold in gradient.}
        \end{figure}

        The optimized output from the HE ansatz with \(d=5\) is shown in Fig. \ref{fig:linear_H3_optimized} for comparison.
        As same as in Fig. \ref{fig:H2_optimized}, the circuit parameters seems to be tuned as to minimize the Hamiltonian at around the middle of the trained points.
        It is as expected because intuitively the state which minimizes the Hamiltonian at the middle point would minimize the cost Eq. (\ref{eq:cost}).
        However, the output at \(r=\) 0.6 fails to achieve even to the HF energy.
    
        Fig. \ref{fig:linear_H3_optimized} (b) shows the depth dependency of the optimized output from the AI ansatz.
        The results are generated with the same method as the previous section.
        The output at the trained points gets close to FCI energy as we increase the depth of the circuit, however, we can observe the complex behavior in deeper circuits, which is not apparent in \(\text{H}_{2}\) problem.
        We suspect that it is due to the more complicated structure of Hamiltonians of this problem.
        The complex behavior is similar to the overfitting problem which is encountered in standard machine learnings.
        In general, overfitting happens when the model parameter, in this case, \(\bm{\theta}\) and \(\bm{\varphi}\), are too many in number with respect to the trained points.
        Several approaches can be taken to address the problem.
        The immediate one is to increase the number of the training points, which has a severe disadvantage in the time required for the optimization.
        The other approach without increasing the optimization time is stopping the optimization before the exact minimum is found.
    
        Considering these, we conducted experiments with a threshold on gradients at each iteration; optimization is stopped when the norm of the gradient goes below the threshold.
        Fig. \ref{fig:linear_H3_optimized} (c) shows the result with such threshold.
        It is apparent that the complex behavior is suppressed with this approach.
        However, this approach reduces the accuracy of the optimized output at the trained points.
        It seems that to recover the accuracy that is lost by this approach, it is necessary to increase the depth of the circuit.
        Therefore we conclude the simultaneous optimization approach does not fit as an application of NISQ devices.

        \subsection{Discussion}
        We suspect the reason why this simultaneous optimization approach does not work is due to the fact that, for fixed (optimized) parameter \(\bm{\theta}^*\) and \(\bm{\varphi}^*\), the change in ansatz state \(\ket{\psi(\bm{x},\bm{\theta}^*,\bm{\varphi}^*}\) with respect to the change in the parameter \(\bm{x}\) is not same as in the exact ground state \(\ket{\psi_g(\bm{x})}\).
        The ansatz \(\ket{\psi(\bm{x},\bm{\theta}^*,\bm{\varphi}^*}\) that outputs approximate ground states with respect to different \(\bm{x}\)'s must have the approximately same gradient with respect to \(\bm{x}\), at least around trained points \(\{\bm{x}_\alpha\}\), as the exact ground state \(\ket{\psi_g(\bm{x})}\).
        This can be a guiding instruction on constructing quantum circuits for VQE.
        Interpolation can be regarded as a way of realizing this concept.
    
    \bibliographystyle{apsrev4-1}

    \end{document}